\documentclass[twocolumn,showpacs,prl]{revtex4}

\usepackage{epsfig}
\usepackage{amsmath}
\usepackage{graphicx}
\usepackage{dcolumn}
\usepackage{amsmath}
\usepackage{latexsym}

\begin{document}

\title{Magnetoresistance of mesoscopic granular ferromagnets }
\author{A.Y. Dokow}
\author{H. Vilchik}
\author{A. Frydman}
\address{The Minerva Center, The Department of Physics, Bar Ilan University, Ramat Gan 52900, Israel}

\begin{abstract}
We have performed  magnetoresistance (MR) measurements of granular
ferromagnets having lateral dimensions smaller than 0.5 $\mu$m and
containing a small number of grains (down to about 100). Compared
to macroscopic samples, these granular samples exhibit unusually
large saturation fields and MR amplitudes. In addition, the
evolution of the magnetoresistance curve as the intergrain
distance decreases is qualitatively different than that of large
samples. We discuss these results and suggest that they reflect a
transition from percolation to quasi single-channel dominated
transport.

\end{abstract}

\date{\today}

\maketitle

\textbf{\centerline{I. INTRODUCTION}}
\bigskip

Insulating granular ferromagnets, i.e. systems of magnetic grains
imbedded in a non-magnetic insulating matrix, exhibit negative
magnetoresistance (MR) curves. At zero magnetic field (H=0) the
resistance is maximal. For applied $\pm$ H the resistance
decreases until, for large enough field, it reaches saturation.
This behavior was observed in a variety of granular samples
\cite{gittelman,milner,yang,honda,sanker,aviad1,aviad2}  and has
been ascribed to spin dependent tunneling between randomly
oriented magnetic moments of the grains \cite{abeles,sheng}.
Applying a magnetic field aligns these moments, causing the
tunneling resistance to decrease. The magnetoresistance is thus
determined by the relative magnetic orientation of pairs of
grains. The MR amplitude (defined as $\Delta R/R
=\frac{R(H\rightarrow\infty)-R(0)}{R(0)})$ of a pair of grains i
and j is given by \cite{slon}:
\begin{equation}
\frac{\Delta R}{R}=\frac{1+P^{2}cos(\alpha)}{1+P^{2}}-1
\end{equation}
where P is the electron polarization and $\alpha$ is the angle
between the moment orientations at zero magnetic field. In the
presence of magnetic field, H, the magnetic orientation of each
grain is governed by two factors. The first is the magnetization
easy axis of the grain due to its anisotropy, and the second is
the external magnetic field. The total magnetic energy of a grain
per unit volume is given by \cite{melikhov}:
\begin{equation}
\frac{W}{V}=\frac{\mu_{0}}{2}M_{S}^{2}\nu
sin^{2}(\theta)-\mu_{0}M_{S}Hcos(\phi)
\end{equation}
where $\mu_{0}$ is magnetic permeability in vacuum, $\nu$ is the
anisotropy coefficient  (of the order of 1), $M_{S}$ is the
saturation magnetization, $\theta$ is the angle between the
magnetic moment orientation and the easy axis of magnetization and
$\phi$ is the angle between the magnetic moment and the applied
magnetic field. Eq. 2 is composed of two energies. The left
component, $W_{I}$, represents the energy due to orientation out
of the easy axis and the right component, $W_{H}$ is due to
alignment of the moment with an external field. The magnetic
orientation of each grain (and therefore the MR amplitude) at zero
temperature is determined by minimizing the energy of equation 2.
It should be noted that ultrasmall grains at finite temperatures
undergo a superparamagnetic transition, thus the thermal energy,
$k_{B}T$, may dominate and overcome the effect of the magnetic
energy.

When dealing with transport properties through a granular
insulator, one has to take into account that not all the grains
participate in the conductance processes since it is a strongly
disordered system. It has long been realized that a percolation
treatment is the proper way to deal with such a system as it
provides much insight into the physics of the conductivity
\cite{pollak, ambegeokar,adkins}. In this approach each pair of
grains i and j is represented by a resistor with resistance
$R_{ij}$ inversely proportional to the hopping probability between
the grains and given by \cite{miller}:
\begin{equation}
R_{ij}=R_{0}exp[\frac{2r_{ij}}{\xi}+\frac{Eij}{k_{B}T}]
\end{equation}
where $r_{ij}$ is the distance between the grains,  $\xi$ is the
localization length representing the decay of the electronic
wavefunction in the insulator and $E_{ij}$ is the energy
difference between the electronic states. For metallic grains at
temperatures of the order of a few K, $E_{ij}$ is related to the
charging energies of the grains which depend on the grain
diameters. In general, smaller grains give rise to larger
$E_{ij}$s. The granular system can be mapped by a resistor network
containing series and parallel current paths. Because the network
contains a wide distribution of resistances (due to the
exponential factors in Eq. 3) the transport is percolative and
governed by a set of critical resistors. These act as "red bonds"
of the percolation network and their conductivity determines the
transport properties of the entire system. Hence, the scale of
inhomogeniety is the percolation radius, $L_{C}$, which can be
viewed as the average distance between critical resistors.

As the lateral dimension of the sample is reduced below $L_{C}$,
the nature of the transport is expected to change dramatically. A
percolation network is no longer a suitable way to treat the
system. Instead, a single current path, or even a single critical
resistor, is expected to dominate the transport, giving rise to
mesoscopic effects and large sample to sample fluctuations.

In this paper we study the MR curves of granular ferromagnets with
sizes smaller than $L_{C}$ ($\sim 0.5 \mu m$) and compare them to
the properties of macroscopic samples. We find that reducing the
size causes a dramatic increase in both MR amplitudes and
saturation fields. In addition, the evolution of the MR curve as a
function of average inter-grain separation is different than that
of large samples. We discuss these results and attribute them to a
crossover from percolation transport to a single dominating
current trajectory, as the sample size is reduced below the
percolation radius.

\bigskip

\textbf{\centerline{I. EXPERIMENTAL}}

\bigskip

The samples described in this paper were discontinuous Ni films
prepared by "quench condensation"
\cite{strongin,bob,goldman,granular rich}, i.e. evaporation of
thin, discontinuous films on substrates that are kept at cold
temperatures and under UHV conditions as described elsewhere
\cite{aviad1,aviad2}. This techniques enables one to perform
electric and magnetic measurements during the sample growth. Thus,
one can study the magneto-transport properties of a \emph{single}
granular Ni sample as a function of film thickness. We note that
the thickness barely changes during the experiment. Adding
$\sim\AA$ to a film of nominal thickness 30 \AA\ is sufficient to
reduce the sample resistance by a few orders of magnitude
\cite{aviad2}. This demonstrates that the decrease in resistance
occurs due to increasing of inter-grain coupling while the grain
sizes remain practically constant during the sample growth
process. Thus the quench condensation method enables one to study
the magneto-transport of granular ferromagnets as a function of
the mean inter-grain distance without thermally cycling the sample
or exposing it to atmospheric conditions.

In order to study small sized granular samples we combined
photo-lithography and atomic force microscope (AFM) fabrication
methods with quench-condensation. First we prepared Ni wires
having width of a few $\mu m$ on a $SiO_{2}$ substrate by
conventional lithography. Next we used an AFM tip (with the z
feedback loop disabled) to cut the wire in two, thus creating two
close electrodes. The nano-space formed between the electrodes
defined the geometry of the measured sample. The properties of the
nano-space were determined by the quality of the AFM's tip, the
force that was applied on the wire and the smoothness of the Ni. A
typical electrode configuration is shown in the inset of fig. 1.
Once a desired geometry was achieved, the substrate was placed on
a quench-condensation probe and a granular film was evaporated
into the gap. Using this technique we were able to prepare samples
with sizes as small as a few tens of nm. Since the grain diameters
are of the order of 10-15 nm \cite{aviad3,aviad4} these samples
contain about 100 grains. We name all samples with lateral
dimensions smaller than $0.5 \mu m$ "mesoscopic samples", while
larger samples are named "macroscopic samples"

\begin{figure}
\centerline{\epsfxsize=3.2in \epsffile{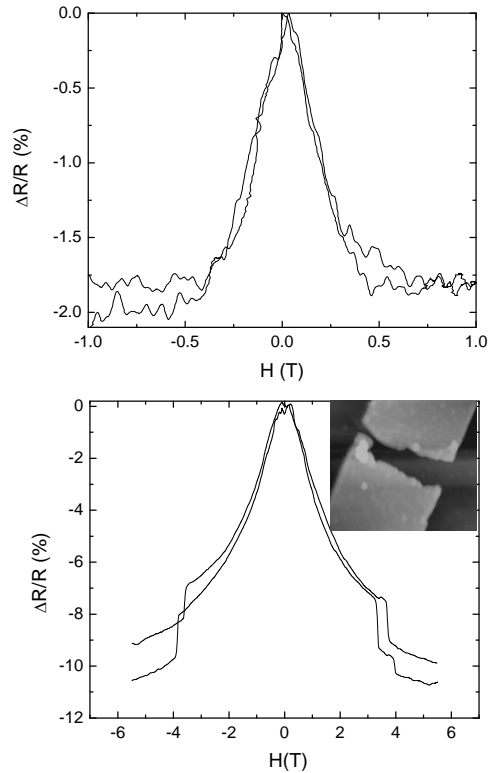}} \vspace{-0.5cm}
\caption{$\Delta R / R$ for a $3mm\ast 3mm$ sample (top panel) and
a $100nm\ast 100nm$ sample (bottom panel). The insert is an AFM
image of the mesoscopic electrode template. A gap with lateral
dimensions of 100nm is cut from a Ni wire having width of $3\mu
m$. } \vspace{-0.4cm}
\end{figure}

\begin{figure}
\centerline{\epsfxsize=3.2in \epsffile{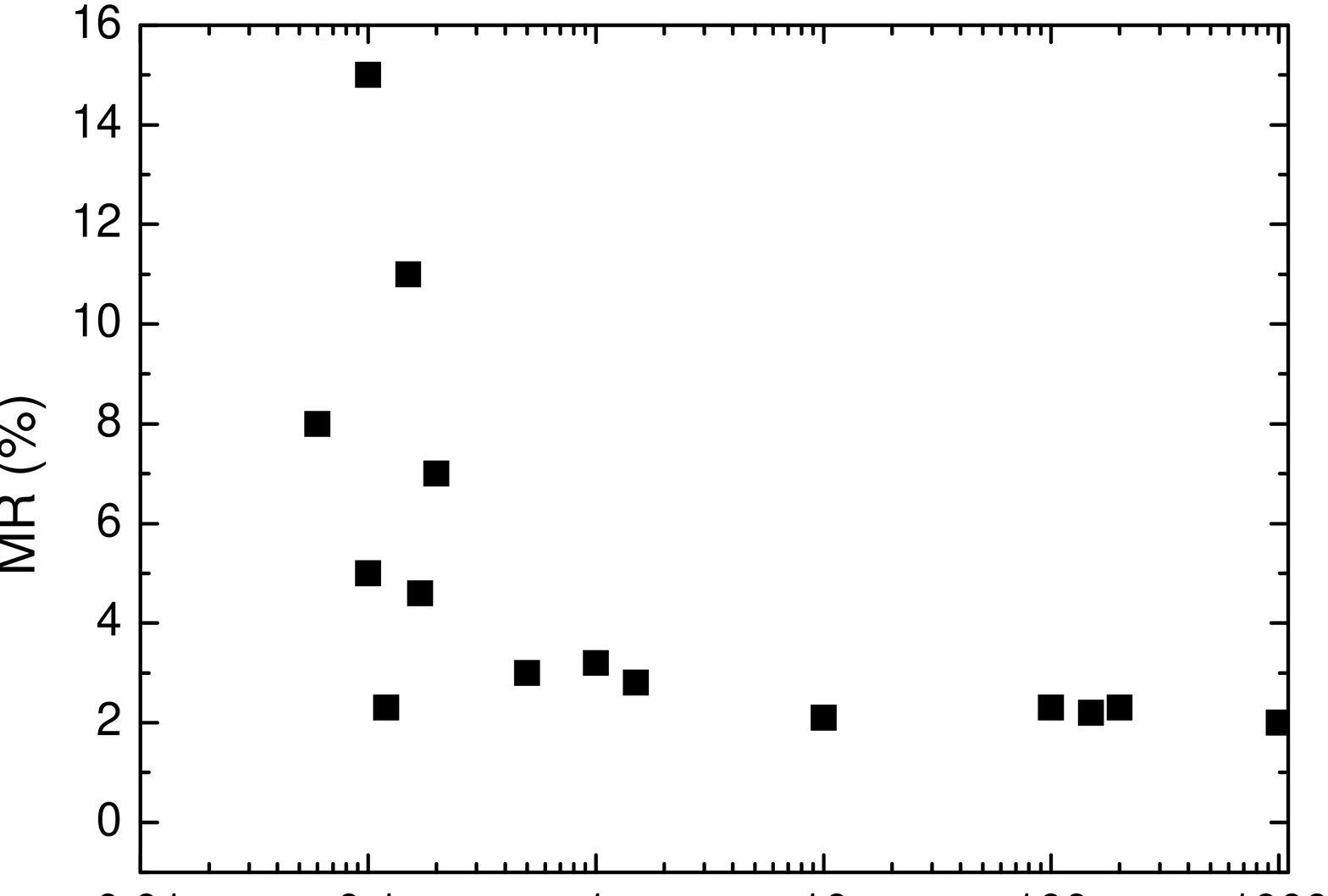}} \vspace{1cm}
\caption{$\Delta R / R$ for different samples having sheet
resistance of $2 M\Omega$ as a function of sample lateral size}
\vspace{-0.4cm}
\end{figure}
\bigskip

\textbf{\centerline{I. RESULTS}}

\bigskip

Magnetoresistance (MR) curves of two samples, one macroscopic and
one mesoscopic, having sheet resistance of 2 $M\Omega$ (nominal
thickness of $\sim 21{\AA}$), are depicted in Fig. 1. The magnetic
field in these experiments was applied perpendicular to the film
plane. The MR amplitude of the macroscopic sample is $2 \%$ and
the saturation field, $H_{S}$, is $\sim$ 0.5T. These values are
typical of all our macroscopic 2D granular samples and are similar
to those obtained by other groups using different preparation
methods for fabrication of insulating granular ferromagnets
\cite{gittelman,milner,yang,honda,sanker}. A saturation field of
0.5T can indeed be expected since it is close to $H = 4\pi M_{s}$,
$M_{s}$ being the saturation magnetization, which is the field
required to align a thin Ni film perpendicular to the substarte
against the shape anisotropy.

Unlike macroscopic samples, the mesoscopic samples exhibited large
sample to sample variations of both MR amplitudes and saturation
fields. $H_{S}$ is always larger than the typical value of 0.5T
and varies between 1T to fields higher than 6T (the largest
available field). $\Delta R / R$ also fluctuates substantially
from sample to sample, however, the average MR amplitude sharply
increases with decreasing sample size as seen in Fig. 2.

As material is added to the sample and the sheet resistance, R,
decreases, several trends are observed in the macroscopic samples
\cite{aviad1,aviad2}. The saturation field, $H_{s}$, remains
constant at a value of $\sim0.5T$ throughout the entire sample
growth process (Fig. 3c). The MR amplitude, on the other hand,
decreases monotonically and smoothly, until, for $R<0.5k\Omega$,
it is suppressed altogether (Fig. 3a). This can be expected since
adding material causes coalescence of grains. For thick enough
layers the sample is simply a continuous Ni film in which no spin
dependent tunneling resistance is expected.
\begin{figure}
\centerline{\epsfxsize=2.5in \epsffile{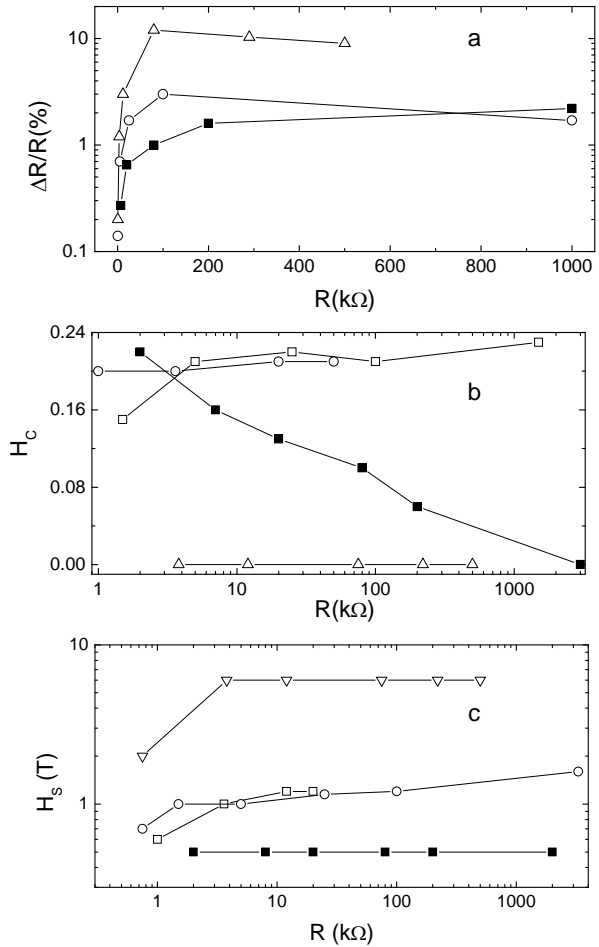}} \vspace{5cm}
\caption{The evolution of $\Delta R / R$ (top panel), the
coercivity $H_{C}$ (middle panel), and the saturation field
$H_{S}$ (bottom panel), as a function of resistance obtained as
material is added to the sample. Full squares are for a
macroscopic sample and open symbols are for a number of mesoscopic
samples. } \vspace{0cm}
\end{figure}

Another observed trend is the splitting of the MR peak. For high
resistance the MR curve is centered at H=0 as seen in Fig. 1. Upon
decreasing resistance, the curve splits into two peaks and a
hysteresis develops in the MR curve \cite{aviad1,aviad2}. The
coercivity, $H_{c}$ grows as a function of decreasing  R and
reaches 0.25T for the lowest resistance in which MR is measurable
(Fig. 3b). This was attributed  to the increase of the effective
grain size as the resistance decreases, thus giving rise to a
transition from a non-hysteretic superparamagnetic sample to a
ferromagnetic sample \cite{aviad1}.

The mesoscopic samples show qualitatively different results. The
MR amplitude does not decrease throughout most of the sample
growth process. On the contrary, $\Delta R / R$ appears to
\emph{increase} initially as material is added to the sample (Fig.
3a). Only for resistances smaller than $50 k\Omega$ a sharp
decrease in the magnetoresistance is observed. A similar effect is
seen for the saturation field, $H_{S}$. It remains constant for
most of the growth and decreases sharply (apparently approaching
the bulk value of $H=4\pi M_{S}\sim 0.5T$) when the resistance
drops below a few $k\Omega$ as seen in Fig. 3c.

Mesoscopic samples differ from macroscopic ones in the evolution
of $H_{C}$ as well. The mesoscopic systems show no development of
hysteresis as a function of resistance. Some of our samples have
no hysteresis for the initial deposition stages, and no hysteresis
is observed even for the lowest measured resistance. Other samples
exhibit a two peak MR curve even for the initial evaporation
stages. In these samples the initial coercivity does not increase
as material is added to the system (see Fig. 3b).

\begin{figure}
\centerline{\epsfxsize=2.6in \epsffile{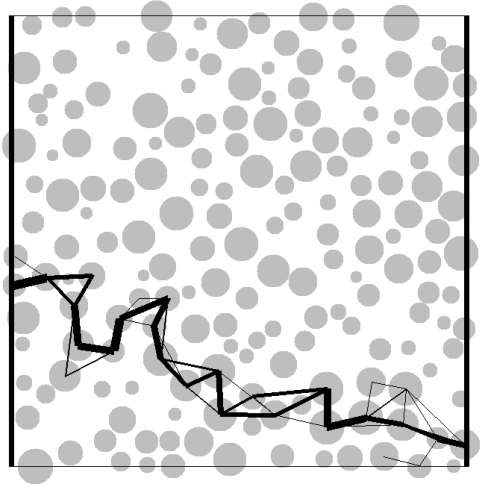}} \vspace{0.5cm}
\caption{Simulation of the current trajectories in a mesoscopic
sample containing 180 grains, showing a single domination current
trajectory. Two leads are connected to the sample on the left and
right edges, with a small d.c. voltage applied between them. The
line width is proportional to the current flow magnitude. }
\vspace{0cm}
\end{figure}
\bigskip

\textbf{\centerline{I. DISCUSSION}}

\bigskip

In order to understand these results we note that our mesoscopic
systems are smaller than the percolation radius, $L_{C}$. We have
performed computer simulations of the MR of our granular Ni
containing a small number of grains \cite{vilchik}. For systems
with less than about 500 grains we find that the transport is
governed by a single dominating channel. The current trajectories
branch out, forming a complex percolating network, only for larger
samples. This corresponds to $L_{C}$ of about 25 grains across,
which, in our samples, is equivalent to a sample size of about 0.5
$\mu m$. For smaller samples the current is forced to flow through
a single dominating chain of grains (see Fig. 4). The mesoscopic
nature of these sample account for the large sample to sample
variations of the MR curves in the small samples. We suggest that
our results can be understood if we assume that the same chain of
grains dominates the transport throughout the sample growth, from
a configuration of electrically isolated grains until the sample
is close to the metallic state. Adding material does not
significantly alter this current path. The resistance decreases
because the inter-grain distance is reduced but the MR amplitude
does not reduce until the grains coalesce, the film becomes
continuous and tunnelling MR is no longer expected. This is very
different to the situation in large samples in which adding
material opens up new channels and modifies the percolation
network.

One consequence of the above model is that the electric current in
small samples may flow through relatively small grains. In
macroscopic samples, hopping through very small grains is
energetically unfavorable due to a large contribution to $E_{ij}$
in Eq. 3. The current can bypass very small grains and choose
larger grains to construct the transport network. This was shown
in reference \cite{aviad2} where the coercivity extracted from MR
measurements was found to be larger than that extracted from
magnetization measurements demonstrating that the percolation
network is constructed of grains larger than the average. In small
samples the situation is different since the transport may be
forced to take place through smaller grains. The fact that only
few grains are present between the electrodes forces the
conducting electrons to hop to small grains with large charging
energy. This may explain the large saturation fields observed in
our mesoscopic samples. The magnetization of grains having
diameters smaller than 10nm was shown to consist of
ferromagnetically aligned core spins and a spin-glass-like surface
layer \cite{kodoma}. Canting of the surface spins introduces
magnetic stiffness of the grain moment since aligning them
requires an extremely large external magnetic field. Another
reason for large $H_{S}$ in small grains is the affect of thermal
fluctuations on the magnetic moment orientation. As shown in Eq.
2, the magnetic energy is proportional to the grain volume. For
small, superparamagnetic grains, the energy due to the external H,
$W_{H}$, has to overcome the thermal energy rather than the energy
due to the easy axis, $W_{I}$. Therefore, the smaller the particle
the larger the affects of surface moments and temperature, giving
rise to large $H_{S}$. As the granular sample size is reduced, the
average grain size participating in the transport decreases. This
may account for the large saturation fields in our mesoscopic
samples. Indeed, we observe a clear correlation between large
saturation fields and small hysteresis in the MR curve (see Fig.
5) reflecting the role played by small grains in causing large
$H_{S}$.

\begin{figure}
\centerline{\epsfxsize=3.2in \epsffile{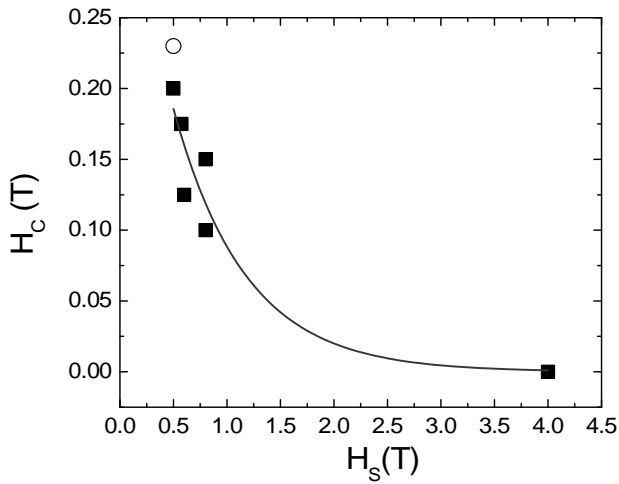}} \vspace{1cm}
\caption{The coercive field, $H_{C}$, determined from the field
position of the MR peak, as a function of the saturation field,
$H_{S}$, for different small samples (solid squares) and a large
sample (empty circle). The solid line is a guide to the eye.  }
\vspace{0cm}
\end{figure}

The behavior of the $H_{C}$ can also be interpreted applying the
above considerations. The crossover from non-hysteretic to
hysteretic MR curves in large samples is attributed to an increase
in the average grain size in the current network.  Adding material
to a granular array causes clustering of the transport network,
adding parallel trajectories which allow the current to flow
through larger grains. $H_{C}$ thus increases as the resistance of
the sample is decreased. In mesoscopic samples this process does
not occur since the current path is relatively fixed throughout
the sample growth process. Hence, the grains that participate in
the transport for high R  are expected to dominate at low R  as
well (though the inter-grain tunneling rate is larger). This
accounts for the constant $H_{C}$ observed in the mesoscopic
samples.

A more subtle issue is that of the MR magnitude. The average MR of
a random array of Ni grains (according to Eq. 1) is expected to be
($\sim\frac{1}{1+P^{2}}-1$). Our simulations \cite{vilchik} show
that for our Ni films this should produce an average $\Delta R/R$
value of about $10\%$ \cite{soulen}. Our macroscopic samples
exhibit significantly smaller values. In addition, Fig. 2 shows
that the smaller the sample the larger is the average MR value.
The reason for this is not clear. Naively, one could expect that a
denser network would emphasize the importance of trajectories with
small mismatch of magnetic moment orientation. Our simulations
show, however, that the average MR amplitude does not depend on
the number of parallel channels. This discrepancy between
experiments and theory impels us to suggest that the MR in the
granular ferromagnets is affected by magneto-static dipole-dipole
interactions which cause alignment of the magnetic moments of the
grains even in the absence of magnetic field. Indeed, we have
observed signs for magnetic interactions in different granular Ni
systems \cite{aviad4}. If ferromagnetic correlations are
significant in the granular sample, they may lead to a reduction
of the measured $\Delta R / R$ relative to the theoretical
expectations. In this case Eq. 2, representing the total magnetic
energy of a grain, i, should contain an additional factor due to
magnetic interaction with a neighbor grain, j:

\begin{equation}
\frac{W_{GG}}{V_{i}}\approx-\mu_{0}M_{S}^{2}\frac{V_{j}}{d_{ij}^{3}}cos(\alpha_{H})
\end{equation}
where $V_{i}$ and $V_{j}$ are the grain volumes and $d_{ij}$ is
the distance between their centers. This energy should be compared
to $W_{I}$ which dominates at H=0. The ratio between the two
energies for two identical grains in contact, having radius a and
volume V, is given by:
\begin{equation}
|\frac{W_{GG}}{W_{I}}|\approx\frac{2}{\nu}\frac{V}{(2a)^{3}}\frac{cos(\alpha_{H})}
{sin^{2}(\theta)}\approx C
\end{equation}
where C is a constant of the order of unity. Thus, the
contribution of the interactions to the magnetic energy is of the
same order of that arising from the easy axis. One can expect that
such interactions will reduce the randomness of the initial
magnetization orientations thus suppressing the MR amplitude. We
note, however, that if the grains are small, the thermal energy
can be strong enough to smear out the interaction effects in a
similar manner to the effect on the easy axis as grains become
superparamagnetic. Therefore, for smaller grains, the MR amplitude
can be larger than that of larger grains. In our mesoscopic
samples, in which the average grain size is smaller, the
interactions will play a smaller role and the MR magnitude may be
larger, as indeed observed in the experiments. This does not
depend on the quench condensed sample resistance since the same
set of grains dominate the transport throughout the sample growth
until the film approaches the metallic phase. In macroscopic
samples, adding material increases the average grain size
participating in the transport as noted above. The importance of
the magnetic interactions (relative to the thermal energy,
$K_{B}T$) thus increases with decreasing resistance. This trend
manifests itself in the reduction of $\Delta R / R$ with film
growth as demonstrated in Fig. 3a.

In summery we point out that reducing the size of any disordered
system, so that it enters the mesoscopic regime, is always
accompanied by rich phenomena. The case where the sample is
ferromagnetic introduces novel effects in addition to the usual
sample to sample fluctuations. These include large MR amplitudes,
large saturation fields and unique coercivity behavior. We suggest
that the experimental findings are indicative of a  transition
from percolative transport to an effective 1D conductance. In the
magnetic case, this enables the detection of the presence of
ultra-small grains which are "invisible" in large samples.
Assuming magnetic inter-grain correlations enables us to account
for the different MR amplitudes measured in macroscopic and
mesoscopic samples. Obviously, A more detailed theoretical
treatment of the issues discussed in this paper is needed. this is
the subject of an ongoing study.

We gratefully acknowledge illuminating discussions with R.
Berkovits and technical help from A. Cohen. This research was
supported by the Israel Science foundation (grant number 326/02).

\end{document}